\documentclass[referee,useAMS,usenatbib]{mn2e}
\usepackage{graphicx}
\usepackage{natbib}
\font\twelvei = cmmi10 scaled\magstep1
       \font\teni = cmmi10 
\font\mbf = cmmib10 scaled\magstep1
       \font\mbfs = cmmib10 \font\mbfss = cmmib10 scaled 833
\font\msybf = cmbsy10 scaled\magstep1
       \font\msybfs = cmbsy10 \font\msybfss = cmbsy10 scaled 833
\def\lsim{\raise0.3ex\hbox{$<$}\kern-0.75em{\lower0.65ex\hbox{$\sim$}}}
\def\gsim{\raise0.3ex\hbox{$>$}\kern-0.75em{\lower0.65ex\hbox{$\sim$}}}

\title[Estimation of the mass of H1743-322 from 2004 outburst]{2004 Outburst of BHC H1743-322: Analysis of spectral and timing properties using the TCAF Solution}

\author[A. Bhattacharjee, I. Banerjee, A. Banerjee, D. Debnath, S. K. Chakrabarti]{\small{Ayan Bhattacharjee\thanks{ayan12@bose.res.in}$^{1}$, Indrani Banerjee\thanks{indrani.banerjee@bose.res.in}$^{1}$, Anuvab Banerjee\thanks{anuvab.banerjee@bose.res.in}$^{1}$,
Dipak Debnath\thanks{dipak@csp.res.in}$^{2}$, Sandip K. Chakrabarti\thanks{chakraba@bose.res.in}$^{1,2}$}\\ 
$^{1}$ S. N. Bose National Centre for Basic Sciences, Block -JD,  Sector -3, Salt Lake, Kolkata 700106, India\\
$^{2}$ Indian Center for Space Physics, 43 Chalantika, Garia St. Road, Kolkata 700084, India
}

\begin{document}

\date{}


\maketitle

\label{firstpage}

\begin{abstract}
The black hole transient H1743-322 exhibited several outbursts with temporal and spectral variability since its discovery 
in 1977. These outbursts occur at a quasi-regular recurrence period of around $0.5-2$ years, since its rediscovery in March 2003. 
We investigate accretion flow dynamics around the Low Mass X-ray Binary H1743-322 during its 2004 outburst using the RXTE/PCA 
archival data. We use Two Component Advective Flow (TCAF) solution to analyse the spectral data. From the fits with TCAF 
solution, we obtain day to day variation of physical accretion rates of Keplerian and sub-Keplerian components, size of the Compton cloud and its other 
properties. Analysis of the spectral properties of the 2004 outburst by keeping fitted normalization to be in a narrow range and 
its timing properties in terms of the presence and absence of QPOs, enable us to constrain the mass of the black hole in a 
range of $10.31 M_{\odot} - 14.07 M_{\odot}$ which is consistent with other estimates reported in the literature.
\end{abstract}


\section{Introduction}

The advent of X-ray astronomy and the launch of Rossi X-Ray Timing Explorer (RXTE) since the last two decades 
have significantly enhanced our understanding about the accretion processes
around compact sources, such as black holes (BHs) and neutron stars (NSs).
Most of these compact sources are in binaries, with the NS or the BH as the primary
which accretes matter from the companion star (secondary) either by Roche-lobe overflow or by capturing the
mass lost from the secondary in the form of winds.
Due to the presence of turbulent viscosity, a part of the gravitational potential energy
lost by the accreting matter is emitted in the form of radiations, chiefly in the soft X-ray domain. Even when 
significant viscosity is absent, stored thermal energy is lost in X-ray domain by inverse Comptonization.
An outbursting Black Hole Candidate (BHC) primarily exhibits four different spectral states,
namely, hard state (HS), hard-intermediate state (HIMS), soft-
intermediate state (SIMS), and soft state (SS) 
(e.g., McClintock \& Remillard, 2006; Nandi et al., 2012; Debnath et al., 2013). 
Simultaneous analysis of the timing properties of these BHCs 
reveal that low frequency Quasi-Periodic Oscillations (QPOs) 
are also evident in the power density spectra (PDS) of these objects
(e.g., Remillard \& McClintock, 2006). Evolution of spectral and temporal 
characteristics of several BHCs during their outbursts have been extensively 
studied (e.g., McClintock \& Remillard, 2006; Nandi et al., 2012).
It has been noted that the various spectral states can be 
related to different branches of the hardness intensity
diagram (HID; Belloni et al., 2005 and  Debnath et al., 2008) or, in a more 
physical hysteresis diagram using accretion rate ratio and X-
ray intensity (ARRID; Mondal et al., 2014; Jana et al., 2016).
The HID or the ARRID shows the objects in different spectral states, generally, in the sequence: HS
$\rightarrow$ HIMS$\rightarrow$  SIMS$\rightarrow$  SS$\rightarrow$  SIMS $\rightarrow$ HIMS 
$\rightarrow$ HS. It is well established that in order to interpret majority 
of black hole spectra two types of spectral components, namely, 
a multi-color blackbody component and a powerlaw component are needed.
The multi-color blackbody component seems to originate from an optically thick, geometrically thin Keplerian flow (Shakura \& Sunyaev, 1973)
while the powerlaw tail of the spectrum is believed to be emanated from 
a ``Compton" cloud (Sunyaev \& Titarchuk, 1980, 1985).
Several theoretical and phenomenological models
ranging from a magnetic corona (Galeev et al., 1979) to a hot gas corona
over the disc (Haardt \& Maraschi, 1993; Zdziarski et al., 2003)
to a two component advective flow (TCAF) solution (Chakrabarti \& Titarchuk, 1995; hereafter CT95)
exist in the literature which attempts to explain the spectrum 
and expound the nature and origin of this ``Compton" cloud.
In this paper, we use TCAF solution to investigate the spectral and timing properties of the source during its 2004 outburst.

The Galactic transient low mass BHXB H1743-322 
is located at $\rm R. A.= 17^h46^m15^s.61$ and $\rm Dec = -32^o 14'00''.6$ (Gursky et al., 1978). 
The discovery of this source goes back to August-September
1977, when Kaluzienski \& Holt (1977) reported its first X-ray activity with the Ariel V All
Sky Monitor. This was subsequently followed by observations from the HEAO I
satellite (Doxsey et al., 1977). Further activities of the source in the
X-ray band 12-180 keV were observed during the 1977-78 outbursts with the HEAO I 
satellite (Cooke et al., 1984). 
Based on the color-color diagram, obtained from the
spectral data of the HEAO I satellite, White \& Marshall
(1984) classified the source to be a potential BHC.
Since its first detection, it remained in the quiescent state till 1984 when
EXOSAT observations reported X-ray activities (Raynolds,
1999) which was subsequently followed by detection of activities by TTM/COMIS instruments on board MirKvant in 1996 (Emelyanov et al., 2000).
On March 21, 2003, the INTEGRAL
satellite detected a bright source named IGR J17464-3213 (Revnivtsev et al., 2003) 
which displayed X-ray activities and later, RXTE confirmed the presence 
of such an activity from the same region in the sky (Markwardt \&
Swank, 2003) validating the source to be H1743-322. 
Since 2003, it exhibited several X-ray activities with quasi-regular intervals of about one to two years.
In order to investigate multi-wavelength properties of the source, 
it was comprehensively monitored in X-rays (Parmar
et al., 2003; Remillard et al., 2006;
McClintock et al., 2009), IR (Steeghs et al., 2003), and in
radio bands (Rupen et al., 2003) during its 2003 outburst. 
McClintock et al. (2009) and Miller-Jones et al. (2012) eventually followed up 
further investigations of the source in the multi-wavelength 
during its 2003 and 2009 outbursts respectively.

The mass of the BHC in H1743-322 has not yet been dynamically measured, although several attempts have been made 
to predict the mass of the BH. Analysing 2003 outburst data, Shaposhnikov \& Titarchuk (2009, ST09) calculated mass of this black hole candidate 
to be $13.3\pm3.2$ $M_\odot$ using their QPO frequency-Photon Index correlation method.
McClintock et al. (2009) estimated its mass to be $ \sim 11 M_\odot$ using their high frequency QPO correlation method.
From the model of high frequency QPOs based on the mass-angular momentum (i.e., spin of the black hole), P\'{e}tri (2008) predicted that its mass should lie in the range of $9-13 M_{\odot}$. 
Based on its spectral and timing properties using two recent outbursts (Molla et al. 2016a, hereafter M16a)
estimated the mass of the BHC to be $M_{BH}=11.21^{+1.65}_{-1.96}$. These authors also used the method of ST09 and narrowed down the range to $11.65\pm0.67M_{\odot}$.
The source is reported to be at a distance of $\rm 8.5 \pm 0.8$ kpc with the 
inclination angle of $\theta \sim 75^o \pm 3^o$. 
Steiner et al. (2012) also constrains to the spin $a$ of the source, 
$-0.3 < a < 0.7$ with a 90\% confidence level.

Recent outbursts of H1743-322 in 2010 and 2011 again showed the characteristic
state transitions (Shaposhnikov and Tomsick, 2010; Shaposhnikov, 2010) as observed in other outburst sources (Nandi et al., 2012).  It was pointed out by Debnath et al. (2010) that depending upon the outburst light
curve profiles, there are mainly two types of outbursting
BHCs: the `fast-rise slow-decay' (FRSD) type and the `slow-rise slow-decay' (SRSD) type. The source H1743-322 belongs to the first category.
Debnath et al. (2013) investigated the average temporal and spectral properties of the object 
using  combined disc black body (DBB) and power law (PL) model using data of 2010 and 2011 outbursts.
Mondal et al. (2014) analysed the RXTE/PCA data during its 2010 outburst using the TCAF 
solution allowing the normalization $N$ of the TCAF model to vary in order to get the best fit. On the other hand, M16a analysed 
the data of H1743-322 using TCAF solution during its 2010 and 2011 outbursts, restricting
the normalization $N$ within a narrow range to estimate the mass of the central object. 
Here, we follow the same procedure of M16a and M16b to analyse the 2004 outburst of H1743-322
which not only enables us to understand the underlying accretion flow dynamics but also allows us to give a fresh
estimate to the mass of the BHC H1743-322. We first determine the average value of normalization by keeping it free within a narrow range. The constant, averaged value of the normalization is then used to refit the spectral data and estimate the mass of the BHC. The 2008b outburst was a `failed outburst' and hence was not analysed.
The 2005 outburst could not be analysed due to lack of data (Coriat et al. 2011). The outburst in 2003 had significant 
radio activity, which is currently being analysed by Nagarkoti et al. (in preparation). From the rest of the cases, 
2004 outburst was the most prominent one, in terms of total flux and duration. 
Hence, we selected this 117 day long 2004 outburst for our analysis. 

Since several outbursts have already been studied one could surmise that analysis of yet another outburst would be of limited use.
If one observed the sequence of outbursts, the one in 2003 took place after about twenty years and after that
there are quasi-regular outbursts, some very small and some moderate. The one in 2003 is very anomalous in the sense that its
intensity was more than five times larger than the next prominent ones, such as those in 2004, 2010, 2011 etc.
and at least 10 to 15 times stronger than several others. This is probably an indication that 2003 outburst could have been triggered by a
non-linear instability and the system is slowly settling and relaxing after subsequent outbursts, probably before going to a long quiescence state again. It is thus no surprise that ST09 estimation of mass of 2003 outburst had a huge error margin while M16a estimate using the same method but 2010 and 2011 data shows a narrower margin. Assuming 2003 outburst is truly anomalous, the first stereotypical outburst is in 2004 and it is important to study this. While fitting the spectra we find that it is highly soft
in much of the time where a standard disk is enough to fit the data.
The region where both spectral components are seen prominently, and both the flow components are
important is in the declining state. So we concentrate only on the declining state of 2004 data.

Judging from the fact that in initial outbursts, the source was going
to very bright soft states and then more relative times have been spent in harder states,
every outburst can be thought to be separately important and combination of the evolution of physical parameters
may lead to the understanding of the long term behaviour of the system. Furthermore, from the light curves
it is easy to see that decay time scales and peak fluxes are different and both of these parameters are
governed by viscosity in the Keplerian flow. The pattern of mass accretion rate
variations are also found to be different from one outburst to the other. Even the accuracy of the estimated mass is vastly different for the same method (M16a).
Thus it is essential that every outburst be studied as accurately as possible.

We organize the paper in the following way: In \S 2 we discuss the salient features of the TCAF solution
and the Propagatory Oscillatory Shock (POS) model.
In \S 3, we discuss observation and methods of data analysis implementing 
HEASARC's HEASoft software package. In \S 4, we present the results obtained from the spectral analysis using the 
TCAF solution and the POS model. Finally, in \S 5 we conclude with a brief discussion summarizing our main findings with some remarks for future work.

\section{Method of Analysis}

\subsection{Two Component Advective Flow (TCAF) Solution}

Prior to the launch of RXTE, Chakrabarti \& Titarchuk (CT95) explored the well-established solution of a transonic flow (see,
Chakrabarti, 1990, Chakrabarti 1996, hereafter C96) and proposed that the accretion flow generally exhibits a two-component behavior, 
namely, a viscous Keplerian flow sandwiched by a weakly viscous sub-Keplerian flow (Fig. 1). 
This solution, popularly known as the Two-Component Advective Flow (TCAF) solution in the literature enunciates that the sub-Keplerian 
halo component, envelops the Keplerian disc and since it requires negligible viscosity to accrete, falls into the BH with a much higher 
radial velocity (Soria et al., 2001; Smith et al., 2002; Wu et al., 2002; Cambier \& Smith, 2013; Tomsick et al., 2014).
The sub-Keplerian flow is advective, can reach supersonic speeds and has angular momentum less than that of a Keplerian distribution. 
Hence, it undergoes a centrifugal pressure supported 
shock transition to become subsonic in between the two sonic points. The complete solutions are worked out in detail in 
Chakrabarti, 1989 (hereafter C89), and C96. 
The centrifugal pressure impedes the flow and as a result a standing or oscillating shock is formed depending on whether the 
Rankine-Hugoniot conditions are satisfied (C89, C96). Thus, the flow puffs up in the vertical direction, and forms the CENtrifugal 
pressure dominated BOundary Layer (CENBOL). This CENBOL acts as the Comptonizing cloud which up-scatters the seed black body photons 
coming from the Keplerian disc.
In the natural hard state of a BHC, this is the only component that is present.
The component near the equatorial plane has a viscosity higher than the critical value and is Keplerian in nature 
having the characteristics of a standard disk. This component is not always present close to the black hole. When the outburst is triggered 
by increase of matter and viscosity, this disk formation is initiated from outside and it moves in on a daily basis, increasing the 
supply of Keplerian matter as well as soft seed photons which are intercepted by the CENBOL and are inverse-Comptonized through repeated
scattering. A typical route of the photon emerging from the Keplerian disc to the observer via CENBOL is shown in Fig. 1.
Initially the hard state is formed when the rates were still low and shock front was hundreds of Schwarzschild radii away and the 
advective (halo) component rate increases first due to its short infall time. Subsequently, the Keplerian rate starts to increase since its
angular momentum is high and it is to be transported by viscosity. Here the object goes to hard intermediate states. The rate continues to rise
and the cooling time scale inside CENBOL starts to be smaller compared to the infall time scale when the condition of oscillation of the
shock front is violated and the QPO seen thus far, becomes sporadic and state becomes soft-intermediate. Finally, if Keplerian matter supply is really high, and the viscosity can transport angular momentum very efficiently, soft photons overwhelmingly cools the CENBOL removing it 
altogether and the soft state is produced. This gradual transformation of the size and shape of the CENBOL is clearly depicted 
in Fig. 2. When the companion turns off the active phase, the process is reversed albeit in a different time scale.
Thus, by giving a clear theoretical origin of the Compton cloud (CENBOL) and 
self-consistently amalgamating the synergy and the inter-conversion of the two components through
viscosity, the TCAF solution provides a clear picture of the entire outburst process and obviates the need of phenomenological
models in the subject. 

For the calculations, all the equations are reduced to their dimensionless forms. The important references can be found in Chakrabarti 1989 (hydrodynamical equations), CT95 (spectral, radiative transfer equations), Debnath et al. (2015a). The last one is the first paper after inclusion of TCAF in XSPEC, where important equations (eq. no. 1 to 5) of TCAF solutions are summarised.

Lengths are measured in units of 
$r_S=2GM_{BH}/c^2$ ($G$ and $c$ being gravitational constant and the velocity of light), 
and the accretion rates are measured in units of Eddington rate 
(also a function of the black hole mass $M_{BH}$). The disc accretion rate ($\dot{m_d}$), 
the halo accretion rate ($\dot{m_h}$), the shock location ($X_s$), the shock compression ratio ($R$) and the mass of the BHC ($M_{BH}$) 
are taken as input parameters and a resulting spectrum is generated. The first four parameters are dependent on the flow properties.
Hence, their time-variation reveals the accretion flow dynamics around the object during an outburst. 
Numerical simulations (Giri \& Chakrabarti 2013)  and spectral studies (Ghosh et. al. 2011) of BHCs reveal that the TCAF solution 
is the most general solution for accreting matter onto a black hole. Self-consistency and stability check of the transonic solution by 
Giri \& Chakrabarti (2013) and Mondal \& Chakrabarti (2013) corroborates that an advective flow will eventually give rise to a TCAF
solution (CT95) when viscous stress near the equatorial plane is substantial. This solution therefore invokes two types of energy 
extraction processes into  a single coherent framework: i) viscous dissipation in the Keplerian component to produce soft X-rays and ii) Inverse Comptonization of these soft photons to produce hard photons by stored thermal energy in the weakly viscous CENBOL.

\begin{figure}
  \centering
\includegraphics[height=7.5cm,width=15.0cm]{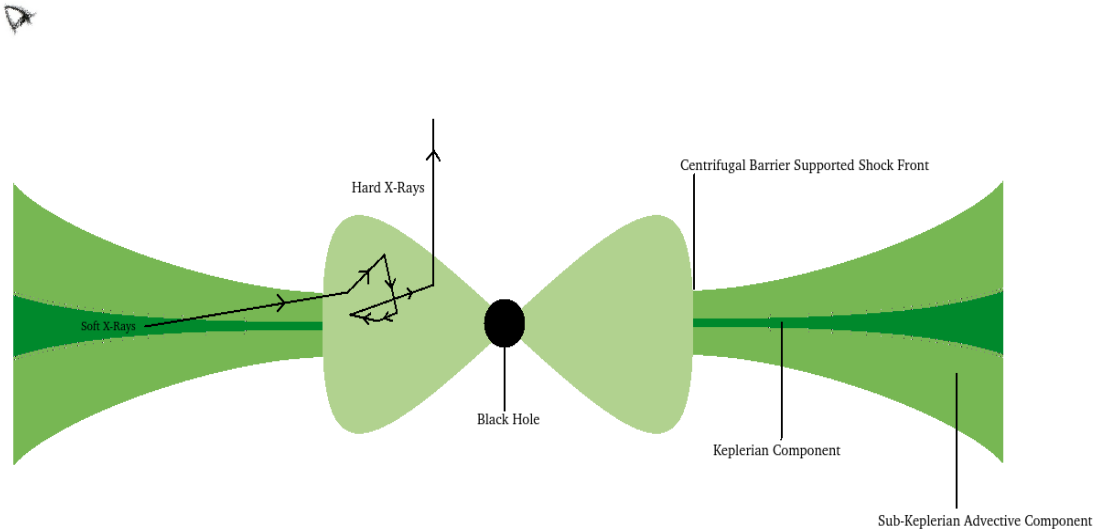}
\vskip-0.5cm
\caption{A schematic diagram of the accretion flow dynamics and radiation processes in a Two Component 
Advective Flow (TCAF) solution. Adapted from CT95.}
\label{lc}
\end{figure}

After obtaining the shape of the overall spectra, suitable model normalization $N$ is used to raise or lower it to match 
the observed spectra. In diskbb plus powerlaw model, the normalization comes only from the disc 
integrated photon number which is used to obtain the inner edge of the truncated disc. 
In TCAF fits, one cannot separate the black body and powerlaw components since the disc 
radiation and its Comptonized spectrum are summed up along with the reflected components in the fits file. The information about 
the inner edge of the truncated disc is in determining the shock location which is also the outer edge of the Compton cloud (CENBOL). 
Since these information are already fed into the grid of the fits file, we only require a constant normalization which is primarily 
mass ($M_{BH}$), distance (D in units of $10$ kpc) and inclination angle 
(i) dependent through a functional relation $N \sim [r_S^2/4 \pi D^2] sin(i)$ (Molla et al. 2016b, hereafter M16b). Error in mass determination would give rise to
error in normalization. Ideally if the CENBOL was lying in a plane, the 
inclination angle would be globally constant, unless the disc is precessing (which we do not assume here).
Also, $M_{BH}$ should not vary over the time scales of observation. Since $M_{BH}\sim T^4$, and the spectral fits are sensitive 
to the temperature $T$, a small error in determination of $T$ gives rise to a significant error in $M_{BH}$. This in turn is 
reflected in the normalization. Moreover, there can be changes in the peak flux with spectral states when the CENBOL 
changes its shape and size self-consistently. The variation of flux is due to the variation of accretion rates 
($\dot{m_d}~and~\dot{m_h}$), shock location ($X_s$) and compression ratio ($R$). In any case, our result is independent of
the exact value of normalization, and our requirement is that it may remain in a narrow range so that we are certain
the fitting routine stays in the same global minimum. The average value, 
for statistical reasons, is taken as the constant value of normalization for the outburst and the entire study is repeated with 
this constant normalization. Under this assumption, if the $M_{BH}$ fluctuates, $D$ and $i$ have to adjust to ensure a constant $N$,
though that does not affect our analysis. For all practical purposes, this $N$ is `fixed' from one outburst to the other as it should be
when precession of the disc is absent.

Recently, the TCAF model (CT95; Chakrabarti, 1997) has been successfully incorporated in HEASARC's spectral analysis
software package XSPEC (Arnaud, 1996) as a local additive table model (Debnath, Chakrabarti \& Mondal 2014; 
Mondal, Debnath \& Chakrabarti 2014; Debnath, Mondal \& Chakrabarti 2015a; Debnath, Molla, Chakrabarti \& Mondal 2015b; 
Jana et al. 2016; Chatterjee et al. 2016; Mondal, Chakrabarti \& Debnath 2016; M16b). 
It accomplishes fitting of the spectral data of several transient BHCs 
(e.g., H 1743-322, GX 339-4, MAXI J1659-152, MAXI J1836-194), during their X-ray outbursts which in turn 
enables us to get a much clearer picture of the accretion flow dynamics in terms of the disc and halo mass accretion rates, 
the location and size of the Comptonizing cloud (here, CENBOL) and the strength of the shock, which in conjunction with 
the size and halo rate gives an idea of the optical depth.

 \begin{figure}
  \centering
\includegraphics[height=7.5cm,width=15.0cm]{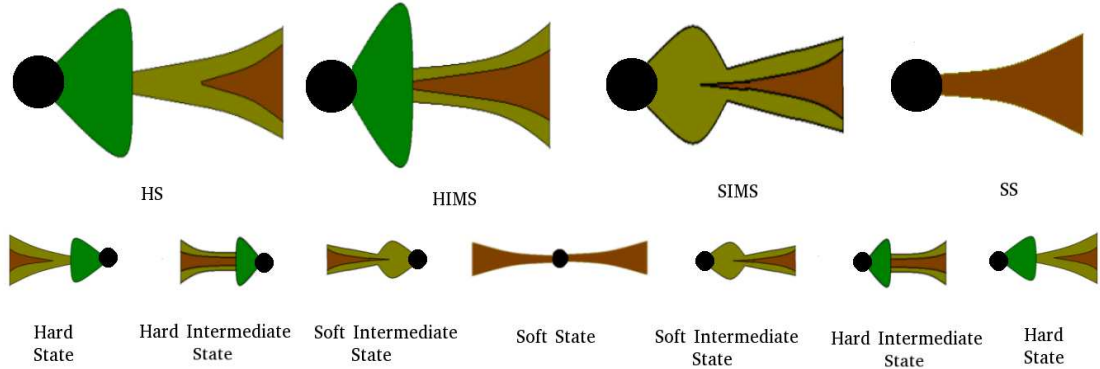}
\vskip-0.5cm
\caption{A schematic diagram showing the evolution of the CENBOL and the Keplerian disc in a Two Component Advective Flow (TCAF) 
solution. At the onset of an outburst, the halo is dominant and the shock front is far away (hard state: HS). 
The Keplerian component moves inward, cooling the CENBOL and making it smaller (hard intermediate:HIMS). For high disc accretion rate, first the 
CENBOL cools and QPOs became sporadic (soft Intermediate state: SIMS) and finally the halo component is
cooled down totally (soft state: SS). The reverse sequence follows when the supply at the outer edge halted.
Adapted from Chakrabarti (2016).}
\label{lc}
\end{figure}

\subsection{Propagatory Oscillatory Shock (POS) model}
Once the Keplerian disc is formed, the CENBOL is cooled down gradually. 
If the cooling timescale of the CENBOL lies within $\sim 50\%$ of the infall time scale from the shock front to the black hole, 
a resonance occurs when the shock or the outer edge of the Compton cloud starts to oscillate (Molteni, Sponholz \& Chakrabarti, 1996; 
Chakrabarti \& Manickam, 2000; Chakrabarti et al. 2015). The oscillation of the shock front leads to the generation of low frequency 
quasi-periodic oscillations (QPOs) in the power density spectra of the light curves. Due to variation of shock conditions with the 
changes in disc and halo accretion rates, the average shock location moves inward (outward) for the rising (declining) 
phase of the outburst (Fig. 2). Simultaneously, the QPO frequencies evolve as well.  

The frequency of oscillation of the shock front, which in turn is related to the QPO frequency ($\nu_{qpo}$) is obtained 
from the inverse of the infall time ($t_{infall}$). If, $c$ is the speed of light, $G$ is the gravitational constant and $M_{BH}$ 
is the mass of the black hole, then the unit of frequency is given by, 
$\nu_{s0}=\frac{c^3}{2GM_{BH}}$. The QPO frequency is then written as,
\begin{center}
$\nu_{qpo}=\nu_{s0}/t_{infall}=\frac{c^3}{2GM_{BH}}\frac{1}{[RX_s(X_s-1)^{1/2}]}$
\end{center}
where, $R$ is the shock compression ratio and $X_s$ is the shock location in units of $r_S$ (Chakrabarti et al. 2008; 
Chakrabarti et al. 2009; Debnath et al. 2010; Nandi et al. 2012; Debnath et al. 2013).
\\
According to the POS model, the shock location varies with time as,
\begin{center}
$X_s(t)=X_s(0)\pm tv_0/r_S$,
\end{center}
where $v_0$ and $X_s(0)$ are the initial location and velocity of the shock. The plus (minus) sign is associated with the shock front 
for declining (rising) phase. This can, in turn, be used to determine the mean radial velocity of the shock front during the outburst 
which further enables us to study the evolution of $X_s$ and hence the evolution of the QPO frequencies.
The POS model has a parametric dependence on the mass $M_{BH}$. Hence, if QPOs are observed in a series of consecutive days, 
then the POS model can be used to fit the variation of the QPO frequency with time. 
The value of mass which gives the best fit to the data will be the mass of the BH.

\section{Observation and Data Analysis}

RXTE/PCA covered the 2004 outburst of H1743-322 spanning from July 11, 2004 (MJD=53197.287) to November 5, 2004 (MJD=53314.749). 
42 observations were recorded by RXTE during the aforementioned period with an average gap of $\sim$ 3 days between consecutive 
observations. We use HEASARC's software package HEASoft, version HEADAS 6.18 and XSPEC version 12.9.0 to carry out our data 
analysis procedure. In order to generate the source and the background ``.pha" files and fit the spectrum exploiting the 
TCAF solution we follow the procedure adopted by Debnath et al. (2013, 2014). For spectral analysis, the Standard2 mode 
Science Data of PCA (FS4a*.gz) were used. Spectra from all the Xenon layers of PCU2 consisting of 128 channels (without 
any binning/grouping of the channels) were extracted for all the observational IDs. Dead-time and pca breakdown correction 
were incorporated in our analysis. We extracted the PCA background by applying the command ``runpcabackest" and by using 
the latest bright-source background model. In order to take care of the South Atlantic Anomaly (SAA) data we incorporated 
the pca saa history file. The task ``pcarsp" was used to create the response files.
For preparing the power density spectra (PDS) all active PCUs were used for a broad energy binned between 0-35 channel data.
The 2.5 - 25 keV PCA spectra of these observation IDs with appropriate background subtraction were fitted with TCAF solution 
based additive model fits file. To accomplish the best fit, a Gaussian line was used to model the iron line emission. 
Throughout the outburst, the hydrogen column density ($N_H$) was kept fixed at $1.6 \times {10^{22}}$ atoms $\rm cm^{-2}$ 
(Capitanio et al. 2009) for absorption model \textit{wabs}. A systematic instrumental error of $1\%$ was assumed. We used 
``err" command to find out $90\%$ confidence error values in model fitted parameters. 

Here we initially fit the entire 117 day long outburst using wabs*(diskbb+PL) model. Next, we fit the last $27$ days 
of the outburst (declining phase) by keeping $\dot{m}_d$, $\dot{m}_h$, $X_s$, and $M_{\rm BH}$ free and normalization within 
the range $10<N<20$ as in Molla et al. (M16a). Next, using the POS model, the velocity of the shock front was found out using 
the parameters as obtained from the previous analysis. The same process of spectral fits using the TCAF model was repeated 
using the average value of normalization obtained from the previous analysis. For this purpose, the model fits file 
(TCAF.fits) was used which uses the theoretical spectra-generating software by varying the five basic input parameters 
in the suitably upgraded CT95 code and is then incorporated in the XSPEC as a local additive model. The version of TCAF 
used for fitting the spectra in the present work is TCAFv0.1.R3.fits used in Debnath et al. (2015a) and 
references therein.

\begin{figure}
  \centering
\includegraphics[scale=0.7]{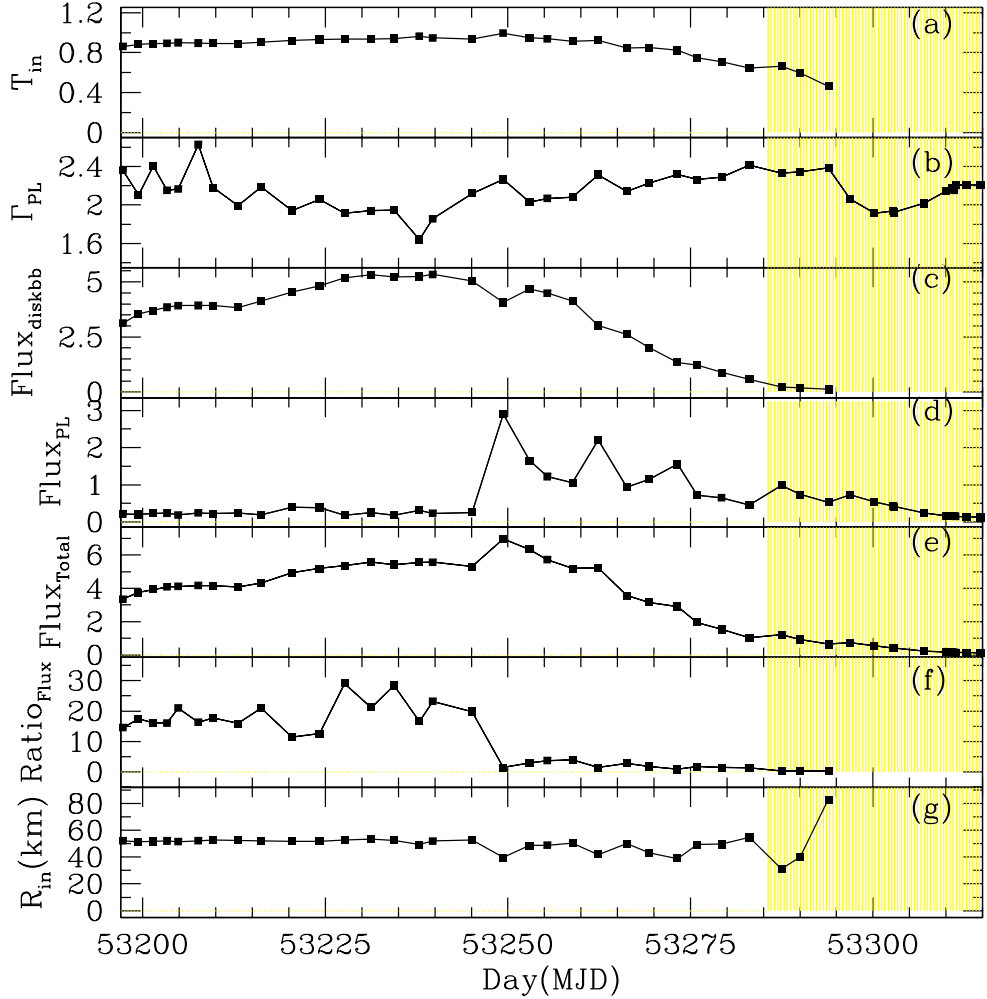}
\caption{Variation of (a) the \textit{diskbb} Temperature ($\rm T_{in}$) in units of keV, 
(b) the \textit{power law} photon index ($\Gamma$), 
(c) the \textit{diskbb}
flux ($\rm Flux_{diskbb}$) in units of $10^{-9}$ ergs $\rm cm^{-2} s^{-1}$,
(d) the \textit{powerlaw}
flux ($\rm Flux_{PL}$) in units of $10^{-9}$ ergs $\rm cm^{-2} s^{-1}$, 
(e) the total flux ($\rm Flux_{Total}$) in units of $10^{-9}$ ergs $\rm cm^{-2} s^{-1}$,
(f) the ratio of \textit{diskbb} flux ($\rm Flux_{diskbb}$) and \textit{powerlaw} flux ($\rm Flux_{PL}$) $\rm, Ratio_{flux}$, and
(g) the inner edge of the disc ($\rm R_{in}$) in units of km, given by the 
diskbb normalization with MJD (days).
All the above variations are in the 2.5-25 keV energy band. The declining phase, that was fitted with TCAF solution, is shaded with yellow (colour online).}
\label{fit0}
\end{figure}

\section{Results}

In this Section, we present the results obtained from the analysis of the data of H1743-322 during the 2004 outburst by the TCAF fits. 

A comparison of RXTE/ASM light curves of H1743-322 between 2003 and 2009 shows that the source predominantly resided 
in the softer states during the 2004 outburst, namely, soft-state (SS) and soft-intermediate state (SIMS), and only towards 
the end of the outburst the source entered in the hard-intermediate state (HIMS) and hard-state (HS) (Fig. \ref{fit0}(e)). 
Capitanio et. al (2005) also reports a similar behavior of the source during this outburst.

Figures \ref{fit0}(a) and \ref{fit0}(b), illustrate variation of \textit{diskbb} temperature $\rm T_{in}$ and the \textit{powerlaw} 
photon index $\Gamma$ with MJD. Figures \ref{fit0}(c) and \ref{fit0}(d) show the variation of the flux
contributed by \textit{diskbb} and \textit{powerlaw}
models respectively, while Fig. \ref{fit0}(e) shows the variation of the total spectral flux. The ratio of \textit{diskbb} flux ($\rm Flux_{diskbb}$) and \textit{powerlaw} flux ($\rm Flux_{PL}$), $\rm(Ratio_{flux}$) is shown in the panel \ref{fit0}(f).
The variation of the inner edge of the disc ($\rm R_{in}$), given by the 
diskbb normalization with MJD is shown in Fig. \ref{fit0}(g).
We calculated the individual flux
contributions for the \textit{diskbb} and the \textit{powerlaw} components 
by using the convolution model ``cflux" once for the \textit{diskbb} and then for the \textit{powerlaw}, 
to fit the spectra in the 2.5-25 keV energy band. 
The analysis with \textit{wabs*(diskbb+powerlaw)} model also reveals that the data for the rising phase was only 
obtained well after the object settled into Soft-Intermediate state. Subsequently, during the rising phase, the peak and 
the initial part of the declining phase of the outburst the underlying accretion flow was primarily governed by a single 
component, i.e., the Keplerian flow. As a result, its spectrum is principally fitted by \textit{diskbb} and the additional 
advective component is not needed. Since in TCAF we are interested to study the interplay between the two components, 
we concentrate on the declining phase during MJD=53287.484 - MJD=53314.749 period.

\subsection{Spectral Data Fitted by TCAF model}
Table \ref{table:tcaf1} illustrates the variation of the physical parameters, namely, the disc accretion rate ($\dot{m}_d$), 
the halo rate ($\dot{m}_h$), the shock location ($X_s$) which gives an indication of the size of the Compton cloud and 
the strength of the shock ($R$) with days (MJD) in the 2.5-25 keV (3-53 channels) energy band during the declining phase 
of the outburst. To explicitly show that the rates are independent, we also plot the ratio ARR(=$\dot{m}_h$/$\dot{m}_d$). The model normalization ($N$) is kept in the range $10-20$. The average value of $N$ ($N_{avg}$) is 
determined and the same procedure is repeated for the constant value of $N=N_{avg}=13.65$,  which we think is a good estimate of the normalization. If we ignore possible precession in the system, contribution from jets which have not been introduced in TCAF, and the possible change in shape with states, this can be treated as true normalization for the system.
The spectral evolution of the rest of the parameters are shown in Table \ref{table:tcaf2}. 
The trends of variation of the flow parameters were found to be similar in both cases. Hence, we only discuss the spectral evolution for $N$ in the range $10-20$. 

\textbf{Spectral Evolution of the Declining Phase and Corresponding TCAF Parameters:}

\begin{figure}
  \centering
\includegraphics[scale=0.8]{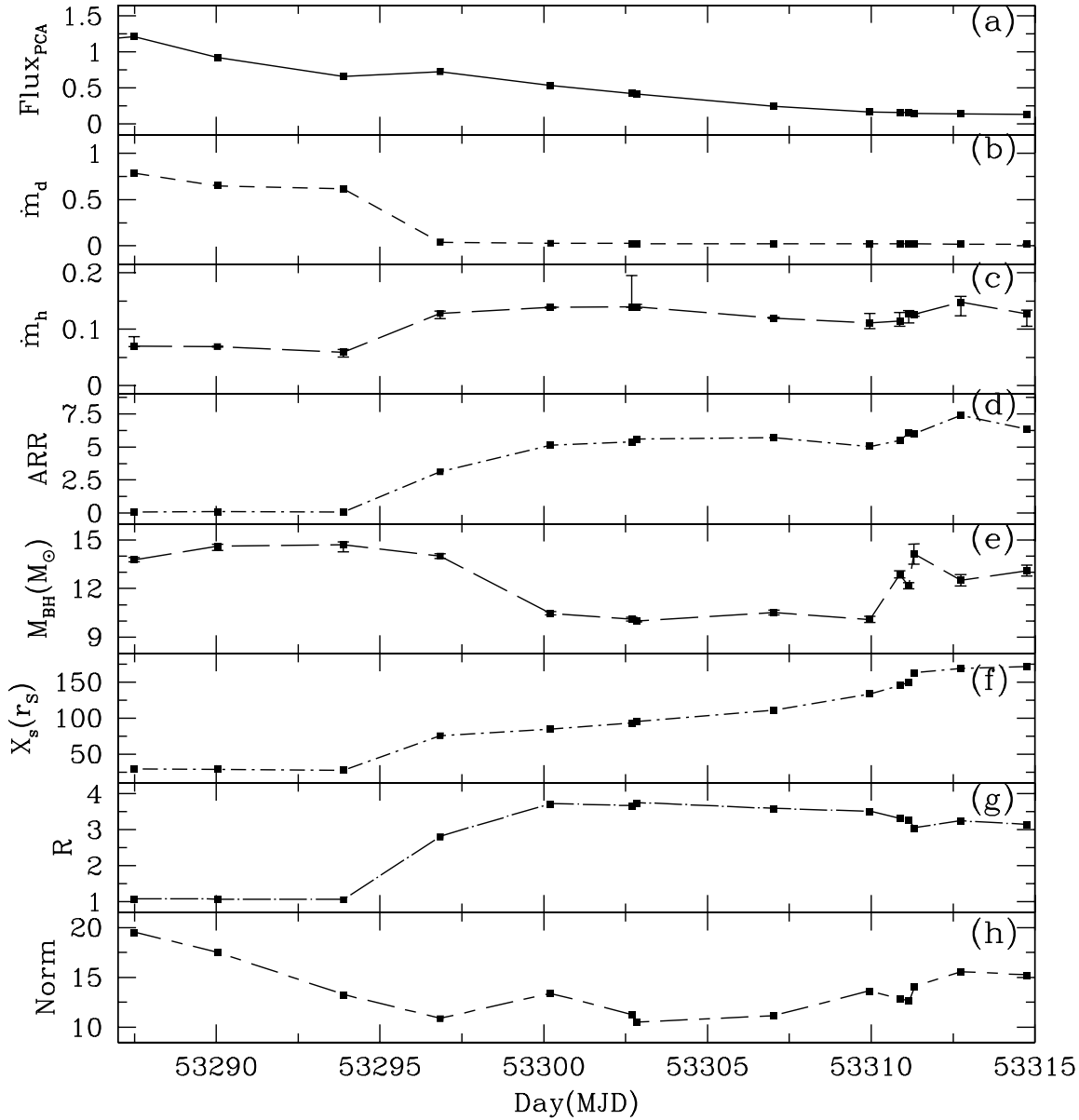}
\caption{Variation of (a) the total PCA flux (in units of $\rm 10^{-9}$ ergs $\rm cm^{-2} s^{-1})$,  
(b) the disc accretion rate $\dot{m}_d$ (in Eddington units),
(c) the sub-Keplerian halo accretion rate  $\dot{m}_h$ (in Eddington units) ,
(d) the accretion rate ratio, ARR($=\dot{m}_h/\dot{m}_d$),
(e) the mass of the black hole (in units of $M_{\odot}$),
(f) the shock location $X_s$ (in units of $r_S$), 
(g) the shock strength $R$, and
(h) the normalization of the TCAF model, with day (MJD). 
Note that the normalization is restricted between $10-20$ in the above fits.
Variation of all the aforementioned quantities are studied in the 2.5-25 keV energy band.
We have added error-bars corresponding to Fig. \ref{fit1}(c) and Fig. \ref{fit1}(e). The remaining 
error-bars are too inconspicuous to be marked.
}
\label{fit1}
\end{figure}

\begin{figure}
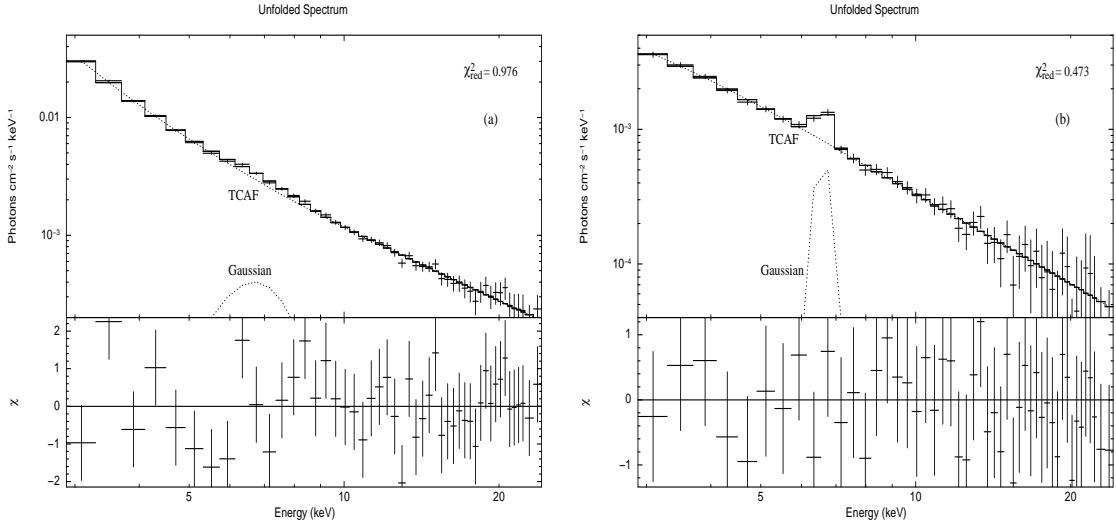

  \centering
  \includegraphics[height=7.5cm,width=7.0cm,angle=-90]{figure5a.eps}
\includegraphics[height=7.5cm,width=7.0cm,angle=-90]{figure5b.eps}
\vskip-0.0cm
\caption{Unfolded spectra with residue of two observations (ID: 90115-01-03-00 (a) and 90115-01-06-00 (b)) for energy 2.5-25.0 keV, 
fitted with \textit{wabs(TCAF+Gaussian)} models.}
\label{fitspec2}
\end{figure}

(i) \textit{Hard-Intermediate State (HIMS)}: For a period of $\sim 13$ days, MJD=53287.484 to 
MJD=53300.208, the object seems to have remained in the Hard-Intermediate state (HIMS) which is evident from
Table \ref{table:tcaf1} and Fig. \ref{fit1}. During this period the PCA flux consistently 
decreases as long as $\dot{m}_d$ is greater than $\dot{m}_h$, from MJD=53287.484 to MJD=53293.882 
and then remains roughly constant (Fig. \ref{fit1} a). On MJD=53296.834, $\dot{m}_h$ becomes greater 
than $\dot{m}_d$ and shock suddenly moves outward from $X_s=27.803 r_S$ to $X_s=75.662 r_S$. During the HIMS 
to HS transition, on MJD=53296.834 and MJD=53300.208, QPOs were observed in the power density spectra (PDS). 
A typical spectra of this state, along with the unfolded models and residue is shown in Fig. \ref{fitspec2}(a). 

(ii) \textit{Hard-state (HS)}: From the data in Table \ref{table:tcaf1} it seems after MJD=53300.208 the source entered 
in the Hard-State (HS) and remained there till the end of the observation. A typical spectra of this state, along with 
the unfolded models and residue is shown in Fig. \ref{fitspec2}b. The PCA flux also went down from this day and 
remained low till the end of the observation which is characteristic of the Hard State. This is evident from Fig. \ref{fit1}a.
No QPOs were observed in the power-density spectrum of the source during these days.
The disc accretion rate which had fallen towards the end of the Hard Intermediate State (HIMS) remains low ($\sim 0.022$) 
while the halo rate remains comparatively higher ($\sim 0.13$). The shock front moves further and further outward with time, 
from $\sim 75~ r_S$ to $\sim 171~ r_S$, in a period of $\sim 14~ \rm days$. The propagation of the shock front not only gives 
the clear picture as promised by the TCAF spectral fits, but also helps us to use another method of verifying the mass of the 
object by an independent method, namely, by the use of Propagatory Oscillatory Shock(POS) model (Chakrabarti et al., 2005, 2008; 
Debnath et al. 2010, 2013; Nandi et al. 2012). The spectral analysis restrict both mass and normalization in a narrow domain. 
We investigate below to check whether the same mass range is obtained from the timing analysis or not.

\begin{figure}
  \centering
\includegraphics[scale=0.8]{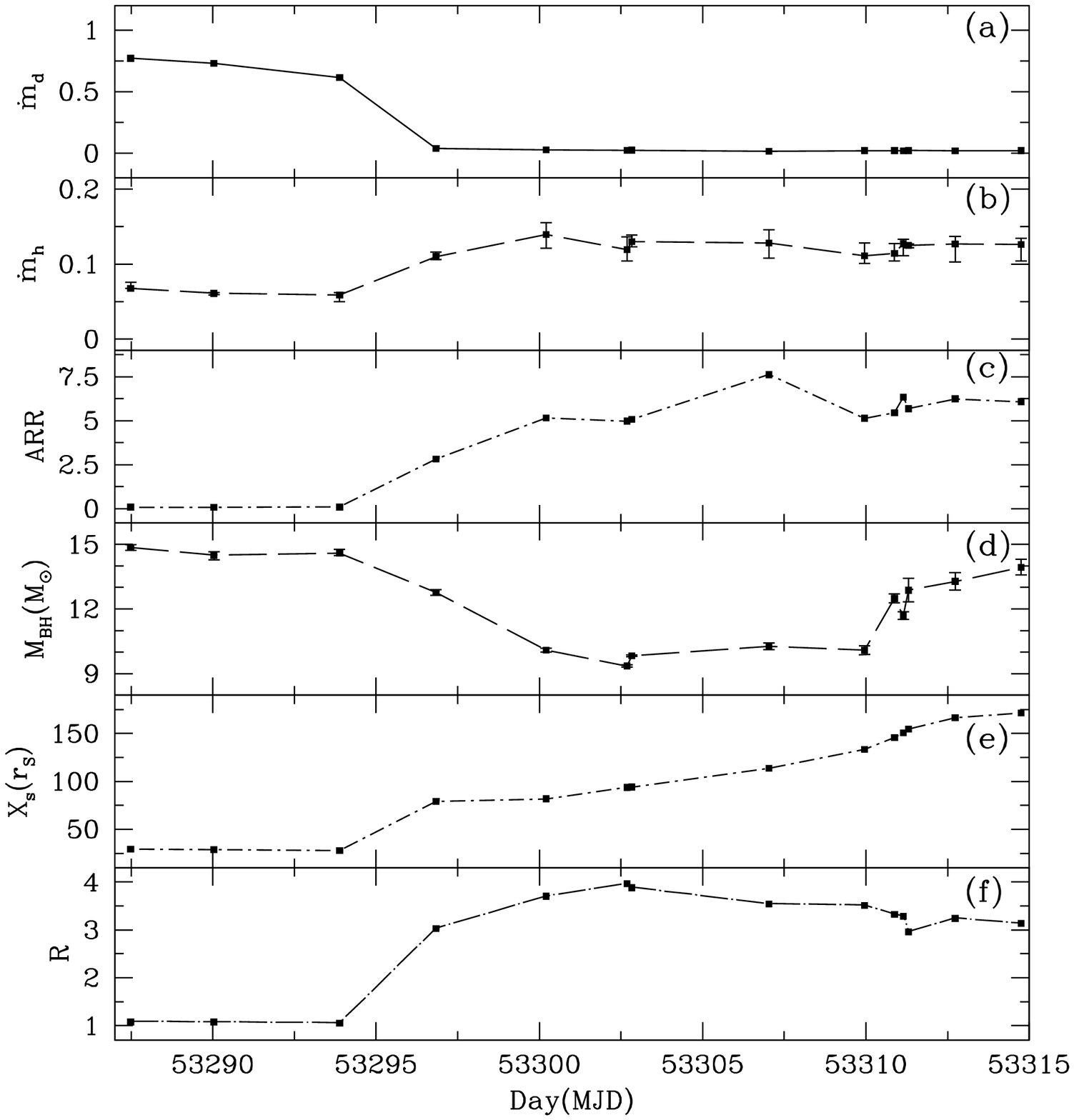}
\caption{Variation of 
(a) the disc accretion rate $\dot{m}_d$ (in Eddington units),
(b) the sub-Keplerian halo accretion rate $\dot{m}_h$ (in Eddington units),
(c) the accretion rate ratio, ARR($=\dot{m}_h/\dot{m}_d$),
(d) the mass of the black hole (in units of $M_{\odot}$),
(e) the shock location $X_s$ (in units of $r_S$),
(f) the shock strength $R$, with day (MJD).
Variation of all the aforementioned quantities are studied in the 2.5-25 keV energy band keeping the 
the normalization of the TCAF model fixed to $N=N_{avg}=13.65$.
We have added error-bars corresponding to Fig. \ref{fit2}(b) and Fig. \ref{fit2}(d). The remaining 
error-bars are too inconspicuous to be marked.
}
\label{fit2}
\end{figure}

The mass was determined to be $\langle M_{BH} \rangle=12.36\pm 1.73~M_{\odot}$, with an average $\langle \chi^2_{red} \rangle = 0.82 \pm 0.20$, and the average normalization is found to be $N_{avg}=13.65\pm 2.49$. The same process, when repeated by using a fixed $N=N_{avg}=13.65$, yielded similar results. These are shown in Fig. \ref{fit2}. 
The mass was determined to be $\langle M_{BH} \rangle=12.19\pm 1.88~M_{\odot}$, with an average $\langle \chi^2_{red} \rangle = 0.79 \pm 0.19$. 
The peak of the Gaussian line profile was close to the same value as that of the previous case. 

\subsection{Correlating spectral and timing properties}

Timing analysis was limited by the lack of observable low-frequency QPOs in the declining phase of the outburst. Out of the 14 data IDs, only 2 days showed prominent QPOs. We use this to determine the mass of the object separately. 
We use the POS model in determining the QPO frequencies from the spectral fit parameters, and use the same equations to obtain 
the value of mass, for which the deviation between the theoretical and observational QPO frequencies are minimum. We also 
determine inward velocity of the shock front for the first day of QPO observation to compare our results with previous works 
(Chakrabarti \& Manickam, 2000; Chakrabarti et al. 2009; Debnath et al. 2013).

Each PDS was analysed with a Lorentzian fit using the ftools commands. The values of the centroid frequency, full-width at 
half-maxima and peak power are obtained with 90\% confidence. From the values of $X_s,\ R,\ M_{BH}$, the oscillation frequency 
of the shock front which is directly related to the frequency of the QPOs can be derived and compared to the frequency obtained 
independently from the analysis of the power density spectrum, $\nu_{QPO}^{PDS}$.

The QPO frequencies obtained from the POS model using the corresponding values of the TCAF parameters taken from Table 1, 
along with their respective systematic errors are given in Table 3, and is denoted by $\nu_{qpo}^{POS}$. The same calculation is 
repeated for the case with constant normalization, using the respective parameters from Table 2 and is reported in Table $4$. 
We found that these two values agree with each other within the error bars.

Next, on MJD=53296.834 and MJD=53300.208, when QPOs were observed, we used the corresponding fit parameters $X_s$ and R 
and varied the mass from 9 to 15 $M_{\odot}$ to calculate the corresponding QPO frequencies, $\nu_{QPO}$. 
The chi-square of the distribution is calculated using $\nu_{QPO}^{PDS}$. The value of the mass for which the chi-square is 
minimum corresponds to $M_{BH}^{POS}$ which is quite close to the value of the mass obtained from the TCAF fits, $M_{BH}$,
on each of the days.

We calculated the velocity of the shock front for the first day (MJD=53296) of QPO observation. The velocity of the 
shock front is $v_0=107.0_{-18.083}^{+18.083}~cm s^{-1}$, which is of the order of values found by Debnath et al. (2013). 
For the constant normalization case, it came out to be $v_0=31.6_{-16.146}^{+16.146}~cm s^{-1}$. This is in the same ball park 
for all the outburst sources GRO J1655-40 (Chakrabarti et al., 2005, 2008), XTE J1550-564 (Chakrabarti et al., 2009), 
GX 339-4 (Debnath et al., 2010; Nandi et al., 2012), H 1743-322 (Debnath et al., 2013) and IGR J17091-3624 (Iyer et al., 2015) 
and is generally thought to be due to the change of pressure in CENBOL due to Compton cooling which drives the shock 
radially (Mondal et al. 2015). The deviation between these two approach may be due to the fact that POS had only two points to fit the evolution of QPOs. With constant normalization, the position of the shock front does not change 
much during the two consecutive days which gives rise to a large systematic error in its measurement which is responsible 
for the discrepancy between the shock velocities in the two cases.


\subsection{Mass estimation using TCAF and POS model fits}

The three different methods used in determining the mass of the black hole, yield masses in the same range, upto the corresponding error-bars. We obtain the average value of mass to be $\langle M_{BH} \rangle=12.36\pm 1.73~M_{\odot}$, for spectral fits with free normalization. The average value of mass is found to be $12.19 \pm 1.88~M_{\odot}$, when the normalization is kept constant at $N=N_{avg}=13.65$. The POS model was applied to both the scenarios for both the observation IDs. In the first case, the masses obtained were $14.011~M_{\odot}$ and $10.479~M_{\odot}$ respectively. For the second case, the masses were $12.156~M_{\odot}$ and $11.125~M_{\odot}$ respectively. All four of these values lie within the range estimated by the spectral fits with constant normalization ($10.31~M_{\odot}-14.07~M_{\odot}$). Incidentally, Molla et al. (M16a) also obtained the average value of normalization $N\sim 15.55$ which lies within the range predicted 
by us. Our result conforms with the previous predictions of Shaposhnikov \& Titarchuk (2009) and McClintock et al. (2009). 
However, both these methods use mass of other BHC as a reference to calculate mass of an unknown BH, but our method gives 
an independent method, where mass can be estimated even from one spectral fit using TCAF solution. We do not require to 
know the mass of other BHCs to estimate mass of an unknown BHC. In Shaposhnikov \& Titarchuk 2009, mass only can be predicted 
if there are sufficient observations in the transition and saturation branches of their QPO frequency-Photon Index correlation 
plot. Similarly, in high frequency correlation method (used in McClintock et al. 2009), mass of those unknown BHCs can be 
predicted, which have shown signature of HFQPOs (so far only 7 BHCs). These sources have shown multiple set (2:3) of HFQPOs.

The iron line emission profile was found to be peaked at around $6.5$ keV for all the fits, with an 
average of $6.57 \pm 0.13$ keV. Fig. \ref{fit1} shows variations of all the TCAF parameters with MJD. 
Thus, we find that the normalization $N$ of the TCAF model does not change over a period of seven years 
i.e., from 2004-2011 indicating that probably the accretion disc is not precessing with significant amplitude.

\begin{table}
\tiny
\vskip0.2cm
{\centerline{\large Table \ref{table:tcaf1}}}
\caption{
TCAF Model Fitted Parameters in 2.5-25 keV energy band for normalization in the range $10-20$. Here, we list the variations 
of disc accretion rate ($\dot{m_d}$) and halo accretion rate ($\dot{m_h}$) in Eddington units, shock location ($X_s$) in 
Schwarzschild radius, shock compression ratio ($R$), mass of the BHC ($M_{BH}$) in $M_{\odot}$ and model normalization ($N$) 
with MJD along with their errors. 
The reduced ($\chi^2$) values for each case is also shown in the last column.}
{\centerline{}}
\begin{center}
\begin{tabular}{c c c c c c c c c c}
\hline
\hline
$\rm Obs. $ & $ \rm Id. $ & $\rm MJD $ & $ \rm \dot{m}_d ~(\dot{M}_{Edd})$ & $ \dot{m}_h~(\dot{M}_{Edd}) $ & $ X_s~(r_S) $ & $ R $ & $ M_{BH}~(M_{\odot}) $ & $ N $ & $\chi^2/dof$ \\
\hline
$\rm 1$ & $\rm X-02-00$ & $\rm 53287.484$ & $ \rm 0.785^{+0.108}_{-0.028}$ & $\rm 0.070^{+0.017}_{-0.001} $ & $\rm 29.790^{+0.019}_{-0.019}$ & $ 1.070^{0+0.001}_{-0.001}$ & $\rm 13.774^{+0.143}_{-0.141}$ & $ \rm 19.496^{+0.165}_{-0.164} $ & $ \rm 37.41/40 $ \\
$\rm 2$ & $\rm X-02-01$ & $\rm 53290.036$ & $ \rm 0.649^{+0.007}_{-0.007}$ & $\rm 0.064^{+0.001}_{-0.001} $ & $\rm 28.862^{+0.023}_{-0.023}$ & $ 1.063^{+0.002}_{-0.001}$ & $\rm 14.611^{+0.127}_{-1.253}$ & $ \rm 17.481^{+0.304}_{-0.303} $ & $ \rm 44.00/40 $ \\
$\rm 3$ & $\rm X-03-00$ & $\rm 53293.882$ & $ \rm 0.617^{+0.055}_{-0.021}$ & $\rm 0.059^{+0.005}_{-0.008} $ & $\rm 27.803^{+0.032}_{-0.032}$ & $ 1.055^{+0.001}_{-0.001}$ & $\rm 14.700^{+0.176}_{-0.438}$ & $ \rm 13.242^{+0.120}_{-0.120} $ & $ \rm 39.04/40 $ \\
$\rm 4$ & $\rm X-03-10$ & $\rm 53296.834$ & $ \rm 0.041^{+0.004}_{-0.004}$ & $\rm 0.128^{+0.004}_{-0.009} $ & $\rm 75.662^{+0.704}_{-0.699}$ & $ 2.804^{+0.010}_{-0.010}$ & $\rm 13.995^{+0.151}_{-0.149}$ & $ \rm 10.856^{+0.102}_{-0.102} $ & $ \rm 40.92/40 $ \\
$\rm 5$ & $\rm X-04-00$ & $\rm 53300.208$ & $ \rm 0.027^{+0.001}_{-0.0003}$ & $\rm 0.139^{+0.001}_{-0.002} $ & $\rm 84.784^{+1.265}_{-1.249}$ & $ 3.717^{+0.046}_{-0.046}$ & $\rm 10.476^{+0.091}_{-0.090}$ & $ \rm 13.355^{+0.156}_{-0.155} $ & $ \rm 41.69/40 $ \\
$\rm 6$ & $\rm X-04-10$ & $\rm 53302.689$ & $ \rm 0.026^{+0.0003}_{-0.005}$ & $\rm 0.140^{+0.055}_{-0.001} $ & $\rm 93.115^{+1.528}_{-1.506}$ & $ 3.660^{+0.158}_{-0.134}$ & $\rm 10.112^{+0.106}_{-0.105}$ & $ \rm 11.228^{+0.171}_{-0.171} $ & $ \rm 35.92/40 $ \\
$\rm 7$ & $\rm X-04-20$ & $\rm 53302.829$ & $ \rm 0.025^{+0.007}_{-0.003}$ & $\rm 0.140^{+0.004}_{-0.003} $ & $\rm 95.118^{+0.657}_{-0.654}$ & $ 3.745^{+0.022}_{-0.021}$ & $\rm 9.981^{+0.044}_{-0.044}$ & $ \rm 10.480^{+0.072}_{-0.071}$ & $ \rm 31.09/40 $ \\

$\rm 8$ & $\rm X-05-00$ & $\rm 53307.032$ & $ \rm 0.021^{+0.005}_{-0.003}$ & $\rm 0.120^{+0.001}_{-0.001} $ & $\rm 114.867^{+0.737}_{-0.733}$ & $ 3.519^{+0.024}_{-0.028}$ & $\rm 10.517^{+0.150}_{-0.140}$ & $ \rm 11.121^{+0.250}_{-0.249} $ & $ \rm 31.58/40 $ \\
$\rm 9$ & $\rm X-05-01$ & $\rm 53309.956$ & $ \rm 0.022^{+0.004}_{-0.004}$ & $\rm 0.111^{+0.017}_{-0.010} $ & $\rm 133.474^{+1.572}_{-1.556}$ & $ 3.505^{+0.032}_{-0.032}$ & $\rm 10.094^{+0.204}_{-0.200}$ & $ \rm 13.625^{+0.238}_{-0.238} $ & $ \rm 17.51/40 $ \\
$\rm 10$ & $\rm X-05-03$ & $\rm 53310.876$ & $ \rm 0.021^{+0.004}_{-0.003}$ & $\rm 0.115^{+0.015}_{-0.010} $ & $\rm 145.277^{+1.281}_{-1.273}$ & $ 3.309^{+0.023}_{-0.023}$ & $\rm 12.873^{+0.229}_{-0.224}$ & $ \rm 12.815^{+0.204}_{-0.203} $ & $ \rm 32.10/40 $ \\
$\rm 11$ & $\rm X-05-10$ & $\rm 53311.146$ & $ \rm 0.021^{+0.004}_{-0.001}$ & $\rm 0.127^{+0.006}_{-0.016} $ & $\rm 149.990^{+1.338}_{-1.329}$ & $ 3.267^{+0.021}_{-0.021}$ & $\rm 12.167^{+0.192}_{-0.189}$ & $ \rm 12.638^{+0.194}_{-0.194} $ & $ \rm 23.74/40 $ \\
$\rm 12$ & $\rm X-05-02$ & $\rm 53311.306$ & $ \rm 0.021^{+0.014}_{-0.004}$ & $\rm 0.126^{+0.002}_{-0.003} $ & $\rm 162.910^{+3.033}_{-2.980}$ & $ 3.041^{+0.053}_{-0.052}$ & $\rm 14.133^{+0.613}_{-0.641}$ & $ \rm 14.016^{+0.550}_{-0.549} $ & $ \rm 33.27/40 $ \\
$\rm 13$ & $\rm X-05-04$ & $\rm 53312.736$ & $ \rm 0.020^{+0.008}_{-0.002}$ & $\rm 0.128^{+0.010}_{-0.024} $ & $\rm 168.780^{+2.106}_{-2.078}$ & $ 3.237^{+0.033}_{-0.033}$ & $\rm 12.504^{+0.365}_{-0.354}$ & $ \rm 15.540^{+0.423}_{-0.422} $ & $ \rm 31.36/40 $ \\
$\rm 14$ & $\rm X-06-00$ & $\rm 53314.749$ & $ \rm 0.020^{+0.007}_{-0.002}$ & $\rm 0.127^{+0.007}_{-0.022} $ & $\rm 171.966^{+2.146}_{-1.900}$ & $ 3.135^{+0.025}_{-0.032}$ & $\rm 13.108^{+0.334}_{-0.325}$ & $ \rm 15.217^{+0.345}_{-0.344} $ & $ \rm 18.94/40 $ \\
\hline
\end{tabular}
\label{table:tcaf1}
\end{center}
\end{table}

\begin{table}
\tiny
\vskip0.2cm
{\centerline{\large Table \ref{table:tcaf2}}}
\caption{
TCAF Model Fitted Parameters in $2.5-25$ keV energy band for normalization, $N=N_{avg}=13.65$.
Here, we list the variations of disc accretion rate ($\dot{m_d}$) and halo accretion rate ($\dot{m_h}$) in Eddington units, shock location ($X_s$) in units of Schwarzschild radius, shock compression ratio ($R$) and mass of the BHC ($M_{BH}$) in $M_{\odot}$ with MJD, along with their errors. 
The reduced ($\chi^2$) values for each case is also shown in the last column.
}

{\centerline{}}
\begin{center}
\begin{tabular}{c c c c c c c c c c}
\hline
\hline
$\rm Obs. $ & $ \rm Id. $ & $\rm MJD $ & $ \rm \dot{m}_d ~(\dot{M}_{Edd})$ & $ \dot{m}_h~(\dot{M}_{Edd}) $ & $ X_s~(r_S) $ & $ R $ & $ M_{BH}~(M_{\odot}) $ & $ N $ & $\chi^2/dof$ \\
\hline
$\rm 1$ & $\rm X-02-00$ & $\rm 53287.484$ & $ \rm 0.771^{+0.065}_{-0.022}$ & $\rm 0.068^{+0.008}_{-0.0003} $ & $\rm 29.153^{+0.027}_{-0.027}$ & $ 1.079^{0+0.001}_{-0.001}$ & $\rm 14.852^{+0.130}_{-0.128}$ & $ \rm 13.65 $ & $ \rm 36.81/41 $ \\
$\rm 2$ & $\rm X-02-01$ & $\rm 53290.036$ & $ \rm 0.730^{+0.020}_{-0.019}$ & $\rm 0.061^{+0.001}_{-0.002} $ & $\rm 28.808^{+0.025}_{-0.025}$ & $ 1.075^{+0.001}_{-0.001}$ & $\rm 14.491^{+0.165}_{-0.216}$ & $ \rm 13.65 $ & $ \rm 44.12/41 $ \\
$\rm 3$ & $\rm X-03-00$ & $\rm 53293.882$ & $ \rm 0.615^{+0.050}_{-0.012}$ & $\rm 0.059^{+0.003}_{-0.009} $ & $\rm 27.768^{+0.032}_{-0.032}$ & $ 1.054^{+0.001}_{-0.001}$ & $\rm 14.594^{+0.167}_{-0.125}$ & $ \rm 13.65 $ & $ \rm 38.71/41 $ \\
$\rm 4$ & $\rm X-03-10$ & $\rm 53296.834$ & $ \rm 0.039^{+0.004}_{-0.004}$ & $\rm 0.110^{+0.006}_{-0.004} $ & $\rm 79.041^{+0.494}_{-0.491}$ & $ 3.026^{+0.016}_{-0.016}$ & $\rm 12.772^{+0.127}_{-0.126}$ & $ \rm 13.65 $ & $ \rm 40.40/41 $ \\
$\rm 5$ & $\rm X-04-00$ & $\rm 53300.208$ & $ \rm 0.027^{+0.003}_{-0.003}$ & $\rm 0.139^{+0.016}_{-0.018} $ & $\rm 81.732^{+1.278}_{-1.260}$ & $ 3.700^{+0.046}_{-0.046}$ & $\rm 10.095^{+0.082}_{-0.081}$ & $ \rm 13.65 $ & $ \rm 41.68/41 $ \\
$\rm 6$ & $\rm X-04-10$ & $\rm 53302.689$ & $ \rm 0.024^{+0.002}_{-0.008}$ & $\rm 0.119^{+0.017}_{-0.015} $ & $\rm 93.727^{+1.872}_{-1.838}$ & $ 3.963^{+0.034}_{-0.033}$ & $\rm  9.356^{+0.061}_{-0.061}$ & $ \rm 13.65 $ & $ \rm 35.63/41 $ \\
$\rm 7$ & $\rm X-04-20$ & $\rm 53302.829$ & $ \rm 0.026^{+0.004}_{-0.001}$ & $\rm 0.130^{+0.009}_{-0.007} $ & $\rm 94.085^{+1.076}_{-1.068}$ & $ 3.883^{+0.086}_{-0.084}$ & $\rm 9.831^{+0.034}_{-0.034}$ & $ \rm 13.65$ & $ \rm 30.09/41 $ \\

$\rm 8$ & $\rm X-05-00$ & $\rm 53307.032$ & $ \rm 0.017^{+0.004}_{-0.001}$ & $\rm 0.128^{+0.018}_{-0.020} $ & $\rm 113.569^{+0.793}_{-0.788}$ & $ 3.538^{+0.014}_{-0.028}$ & $\rm 10.268^{+0.153}_{-0.151}$ & $ \rm 13.65 $ & $ \rm 31.19/41 $ \\
$\rm 9$ & $\rm X-05-01$ & $\rm 53309.956$ & $ \rm 0.022^{+0.004}_{-0.004}$ & $\rm 0.111^{+0.017}_{-0.010} $ & $\rm 133.302^{+1.573}_{-1.557}$ & $ 3.510^{+0.032}_{-0.032}$ & $\rm 10.082^{+0.204}_{-0.199}$ & $ \rm 13.65 $ & $ \rm 17.51/41 $ \\
$\rm 10$ & $\rm X-05-03$ & $\rm 53310.876$ & $ \rm 0.021^{+0.004}_{-0.003}$ & $\rm 0.114^{+0.013}_{-0.010} $ & $\rm 145.746^{+1.276}_{-1.270}$ & $ 3.321^{+0.023}_{-0.023}$ & $\rm 12.488^{+0.208}_{-0.205}$ & $ \rm 13.65 $ & $ \rm 32.10/41 $ \\
$\rm 11$ & $\rm X-05-10$ & $\rm 53311.146$ & $ \rm 0.020^{+0.004}_{-0.001}$ & $\rm 0.127^{+0.006}_{-0.016} $ & $\rm 150.742^{+1.335}_{-1.326}$ & $ 3.281^{+0.021}_{-0.021}$ & $\rm 11.700^{+0.177}_{-0.174}$ & $ \rm 13.65 $ & $ \rm 23.75/41 $ \\
$\rm 12$ & $\rm X-05-02$ & $\rm 53311.306$ & $ \rm 0.022^{+0.001}_{-0.001}$ & $\rm 0.125^{+0.002}_{-0.003} $ & $\rm 154.643^{+3.178}_{-3.117}$ & $ 2.957^{+0.050}_{-0.050}$ & $\rm 12.867^{+0.562}_{-0.538}$ & $ \rm 13.65 $ & $ \rm 33.28/41 $ \\
$\rm 13$ & $\rm X-05-04$ & $\rm 53312.736$ & $ \rm 0.020^{+0.008}_{-0.002}$ & $\rm 0.127^{+0.010}_{-0.024} $ & $\rm 166.567^{+2.131}_{-2.107}$ & $ 3.239^{+0.034}_{-0.033}$ & $\rm 13.282^{+0.408}_{-0.395}$ & $ \rm 13.65 $ & $ \rm 31.35/41 $ \\
$\rm 14$ & $\rm X-06-00$ & $\rm 53314.749$ & $ \rm 0.021^{+0.008}_{-0.002}$ & $\rm 0.126^{+0.008}_{-0.022} $ & $\rm 171.430^{+2.157}_{-1.788}$ & $ 3.135^{+0.025}_{-0.032}$ & $\rm 13.935^{+0.374}_{-0.364}$ & $ \rm 13.65 $ & $ \rm 18.94/41 $ \\
\hline
\end{tabular}

\label{table:tcaf2}
\end{center}
\end{table}

\begin{table}
\tiny
\vskip0.2cm
{\centerline{\large Table \ref{table:tcaf3}}}
{\centerline{Comparison of QPO frequencies obtained from theoretical predictions and actual fits when
$N$ is in the range $10-20$.}}
{\centerline{}}
\begin{center}
 \begin{tabular}{c c c c c c c c} 
 Obs ID. & MJD & $M_{BH}~(M_{\odot})$ & $X_s~(r_S)$ & R & $\nu_{qpo}^{POS}$(Hz) & $\nu_{qpo}^{PDS}$(Hz) & $M_{BH}^{POS}~(M_{\odot})$ \\  
 \hline
 90115-01-03-10 & 53296.834 & 13.995$_{-0.149}^{+0.151}$ & 75.662$_{-0.699}^{+0.704}$  & 2.804$_{-0.010}^{+0.010}$ & 3.957$_{-0.112}^{+0.112}$ & 3.952$_{-0.166}^{+0.157}$ & 14.011$_{-0.547}^{+0.602}$ \\  
 90115-01-04-00 & 53300.208 & 10.476$_{-0.090}^{+0.091}$ & 84.784$_{-1.249}^{+1.265}$ & 3.717$_{-0.046}^{+0.046}$ & 3.359$_{-0.145}^{+0.145}$ & 3.358$_{-0.144}^{+0.208}$ & 10.479$_{-0.620}^{+0.460}$\\ 
 \hline
\end{tabular}
\caption{The spectral parameters from TCAF fits $M_{BH}$, $X_s$ and $R$ are used in the formula obtained from POS for the determination of QPO frequency. The value is listed as $\nu_{qpo}^{POS}$ within errorbars. The mass $M_{BH}$ is then tuned further to reduce the difference between POS prediction and observed QPO $\nu_{qpo}^{PDS}$. The corresponding values is noted as $M_{BH}^{POS}$.}
\label{table:tcaf3}
\end{center}
\end {table}

\begin{table}
\tiny
\vskip0.2cm
{\centerline{\large Table \ref{table:tcaf4}}}
{\centerline{Comparison of QPO frequencies obtained from theoretical predictions and actual fits keeping 
$N=N_{avg}=13.65$.}}
{\centerline{}}
\begin{center}
 \begin{tabular}{c c c c c c c c}
 Obs ID. & MJD & $M_{BH}~(M_{\odot})$ & $X_s~(r_S)$ & R & $\nu_{qpo}^{POS}$(Hz) & $\nu_{qpo}^{PDS}$(Hz) & $M_{BH}^{POS}~(M_{\odot})$ \\  
 \hline
 90115-01-03-10 & 53296.834 & 12.772$_{-0.126}^{+0.127}$ & 79.041$_{-0.491}^{+0.494}$  & 3.026$_{-0.016}^{+0.016}$ & 3.758$_{-0.092}^{+0.092}$ & 3.952$_{-0.166}^{+0.157}$ & 12.156$_{-0.475}^{+0.521}$ \\
 90115-01-04-00 & 53300.208 & 10.095$_{-0.081}^{+0.082}$ & 81.732$_{-1.260}^{+1.278}$ & 3.700$_{-0.046}^{+0.046}$ & 3.697$_{-0.162}^{+0.162}$ & 3.358$_{-0.144}^{+0.208}$ & 11.125$_{-0.656}^{+0.487}$ \\
 \hline
\end{tabular}
\caption{Spectral parameters from TCAF fits $M_{BH}$, $X_s$ and $R$ are used in POS model for the determination of QPO frequency. The value is listed as $\nu_{qpo}^{POS}$ within errors bars. The mass $M_{BH}$ is then tuned further to reduce differences between POS prediction 
and observed QPO $\nu_{qpo}^{PDS}$. The corresponding value is noted as $M_{BH}^{POS}$.}
\label{table:tcaf4}
\end{center}
\end {table}

\section{Discussions and Conclusions} 

In this paper, we investigate evolution of spectral properties of the Galactic 
BHC H1743-322 during the declining phase of the 2004
outburst to study the accretion flow dynamics and to extract the mass of the
BH independently from each observation. This was the first `normal' outburst after about 20 years. However,
data during hard and hard-intermediate states in the rising phase is missing and 
since then it was mostly in soft states except towards the end of the declining phase. 
Hence we concentrate our study only in this end phase
as we expect interplay between the two flow components of TCAF models.
We successfully addressed the evolution of accretion rates of both disc and halo, shock location 
(which represents the size of the Compton cloud) and the compression ratio to have a clear understanding of the outburst.
It is important to note here that the fits obtained by spectral models, such as \textit{diskbb + power law}, does not provide 
any clue about the mass of the object, neither does it explain the accretion flow dynamics around the BHC. 
TCAF solution, on the other hand, gives an independent estimate of $M_{BH}$ from every single observation. 

Molla et al. (M16a) obtained the mass of the black hole in the range $11.2^{+1.66}_{-1.95} M_{\odot}$.
Further, they estimated the mass of the BH with other methods such as 
Shaposhnikov \& Titarchuk 2007 (ST07) using the  Photon Index-QPO frequency correlation technique. The measured mass of the black hole was obtained as $11.61\pm0.62M_{\odot}$. Shaposhnikov \& Titarchuk (2009, ST09) predicted the mass of the black hole to be $13.3\pm3.2M_{\odot}$ using the correlation between their spectral and timing properties while McClintock et al. (2009), estimated the mass $\sim 11.0 M_{\odot}$ using their high frequency QPO correlation method. From the model of high
frequency QPOs based on the spin of the black hole P\'{e}tri (2008) obtained the mass of the BHC in the range of $9-13 M_{\odot}$. 
We have obtained the mass of the black hole in the range  $10.31 M_\odot - 14.07 M_\odot$ which agrees well with these 
previous measurements.

The PCA spectra of the object clearly show that most of the flux is contributed from the soft photons emitted by the disc. 
The flux emitted from the disc is $\sim T^4$ and draws its energy from the loss in gravitational potential energy $\sim \frac{GM_{BH}}{r}$. 
Thus, roughly the spectra obtained by the best fit with TCAF has an intrinsic dependence of $T$ 
on $M_{BH}$ as $M_{BH}\sim T^4$. The spectra at low energy, where the blackbody radiation from the disc is dominating, 
is limited by the resolution of RXTE/PCA. Any error in the measurement, combined with the error in 
the fit parameters which depend on $T$, leads to a significant proportional error 
in the determination of $M_{BH}$. Despite that, the mass is found to lie in a narrow range 
which conforms and restricts further, the previous findings by M16a and P\'{e}tri et al. (2008).

We assumed that the mass, distance and inclination angle of the object are constant such that
the projected area of the disc along the line of sight does not change keeping the model normalization
more or less unchanged provided the instrument response function and the absorption from 
intervening medium are determined correctly which may affect Normalization also. Furthermore, we assumed that 
there was no contribution to X-rays from the jet. 
However, since mass, distance or inclination angle are not accurately known and the 
RXTE resolution is low to generate accurate spectra, we allowed to 
vary both mass and normalization within a narrow range. We also used the average normalization and repeated our analysis.
We found reasonably good $\chi^2$ values as evident from table \ref{table:tcaf1} and table \ref{table:tcaf2}. The average reduced $chi^2$ value is $\langle \chi^2_{red} \rangle = 0.82 \pm 0.20$ for $10<N<20$ and $\langle \chi^2_{red} \rangle = 0.79 \pm 0.19$ for $N=N_{avg}=13.65$. We believe that unless the system parameters (most importantly inclination angle for a precessing disc) change, this Normalization may be used to analyze
subsequent outbursts.

The low-frequency of QPO derived from the power density spectrum fitted parameters was found        
to be in agreement with the one obtained from the POS model if the systematic error is considered.
The velocity of the shock front $v_s$ using the POS model was found to be similar to those 
with the previous work by Debnath et al. (2013). 

The masses obtained by the POS model for both the cases with free and constant normalization agree with the range ($10.31~M_{\odot}-14.07~M_{\odot}$) obtained by the spectral fits with a constant, averaged normalization of $N=N_{avg}=13.65$. The POS model provides a secondary verification to our method and reflects upon the general consistency of our approach. We thus conclude from our analysis that the mass of the BHC is in the range $10.31~M_{\odot}-14.07~M_{\odot}$.

TCAF solves the radiative transfer equation for a steady state two component flow to discuss about the 
spectral properties. The day-to-day evolution of the spectra are, thus, explained in terms of the 
variation of physical parameters which are used in TCAF. This allows us to have a fresh estimate of the 
mass of the black hole, independent of any other observations. The POS model predicts values of QPO frequencies from the spectral 
parameters, and we find that observed QPOs are of similar values. Thus TCAF self-consistently 
puts spectral and timing properties under a common framework. The derived mass is well 
within the range estimated by earlier workers using very different observational data, and model. Presently,
TCAF does not incorporate effects of magnetic fields, spin of the black 
hole or line emissions. These effects are being incorporated and results would be reported elsewhere.

\end{document}